\documentstyle[11pt,newpasp,twoside,epsf]{article}
\def\edcomment#1{\iffalse\marginpar{\raggedright\sl#1\/}\else\relax\fi}
\reversemarginpar

\def\VEL{\:{\rm km\:s^{-1}}}

\begin{document}
\newcommand{\MSOL}{\mbox{$\:M_{\sun}$}}

\newcommand{\EXPN}[2]{\mbox{$#1\times 10^{#2}$}}
\newcommand{\EXPU}[3]{\mbox{\rm $#1 \times 10^{#2} \rm\:#3$}}  
\newcommand{\POW}[2]{\mbox{$\rm10^{#1}\rm\:#2$}}

\title{Modelling the Spectral Signatures of Accretion Disk Winds in Cataclysmic Variables}

\author{Knox S. Long}

\affil{Space Telescope Science Institute, 3700 San Martin Drive,
Baltimore, MD 21218, U. S. A.}

\author{Christian Knigge}

\affil{Dept. of Physics and Astronomy, University of Southampton,
Southampton SO17 1BJ, UK}

\begin{abstract}
Bipolar outflows are known to be present in many disk-accreting
astrophysical systems.  In disk-dominated cataclysmic variables,
these outflows are responsible for most of the features in UV and
FUV spectra.  However, there have been very few attempts to model
the features that appear in the spectra of disk-accreting
cataclysmic variables quantitatively. The modelling that has been
attempted has been concentrated almost entirely on explaining the
shape of C IV.

Here we describe a new hybrid Monte Carlo/Sobolev code that allows
a synthesis of the complete UV spectrum of a disk-dominated
cataclysmic variable.  A large range of azimuthally-symmetric wind
geometries can be modelled.  Changes in line shape in eclipsing
systems can also be studied.  Features in the synthesized spectra
include not only well-known resonance lines of O VI, NV, Si IV,
and C IV, but, with an appropriate choice of mass loss rate and
wind geometry, many of the lines originating from excited lower
states that are observed in HUT, FUSE, and ORFEUS spectra.  The
line profiles of C IV resemble those calculated by previous
workers, when an identical geometry is assumed.  We compare the
synthesized spectra to HUT spectrum of Z Cam, showing that in this
case a reasonably good ``fit'' to the spectrum can be obtained.
\end{abstract}

\section{Introduction}

The evidence that dwarf novae and nova-like cataclysmic variables
(CVs) have winds is overwhelming.  The clearest cases are those in
which C IV, and, less often, Si~IV, N~V, O~VI and Ly$\alpha$ have
P Cygni-like profiles. A good example of wind features in the
nova-like IX Vel was shown by Hartley, Drew \& Long (2001) at this
conference. In systems where emission wings are not observed, the
line centroids of the strong resonance lines are commonly
blue-shifted, suggesting that most of the absorption is being
caused by material moving along the line of sight toward the
observer. Blue edge velocities as large as 3000 $\VEL$ are
observed.

Qualitatively, the features in the spectra are understood.
Absorption predominates when the ion associated with a radiative
transition lies mainly along the line of sight to the UV-emitting
portion of the disk, while emission features arise from photons
scattered into our line of sight from beyond the light cylinder to
the disk. This explains why the emission features are more
prevalent in highly inclined systems.

There have been several attempts to reproduce the shape of the C
IV $\lambda$1550 doublet, beginning with Drew (1987) and Mauche \&
Raymond (1987) who carried out Sobolev radiative transfer
calculations in a spherical wind with kinematic prescriptions for
the wind velocity based on that used for O star winds.  Although
these calculations showed that one could produce reasonable P
Cygni profiles with a spherical wind as long as limb-darkening of
the disk was taken into account, observational evidence suggested
that the winds could be better described as bipolar, or
bi-conical, flows arising from the disk. As a result, Shlosman \&
Vitello (1993) and Knigge, Drew \& Wood (1995) developed Sobolev
and Monte Carlo codes, respectively, to calculate C~IV profiles
for bipolar winds arising from the disk.  They used different,
though qualitatively similar, prescriptions to describe the
velocity and density of the wind as a function of position above
the disk. In addition to an overall mass loss rate $
\dot{m}_{wind}$, both provide ways of parameterizing the mass loss
rate as a function of radius in the disk.  Both allow ways to
describe a narrowly collimated wind or one with a wide opening
angle above the disk. Both assume that the wind proceeds along
stream lines, and that the final velocity of the streamline is a
multiple of the escape velocity at the footpoint of the stream
line.  In the azimuthal ($\theta$) direction, the wind law
reflects angular momentum conservation. In the poloidal ($\rho$-z)
direction, both assume velocity laws similar those developed by
Caster \& Lamers (1979) for O star winds, involving  a scale
height and an acceleration parameter.

However, there have been almost no attempts to model
quantitatively other UV lines in the spectra of dwarf novae and
nova-like variables with bi-conical flows, and as a result, even
the origin of many of the features seen in spectra obtained with
HST, HUT, and FUSE are not well established.  Model spectra of
disks produced by summing appropriately weighted stellar
atmospheres can reproduce the general shape of the continuum over
the UV spectral range. But these models do not match the shapes of
features very well (often due to the fact that the absorption
features have relatively narrow components, which are not
reproduced by the rapidly rotating sections of the disk that
dominate UV emission).

Knigge et al.\ (1997) did attempt to model wind profiles in a HUT
spectrum of Z Cam with their code. They (we) showed that the
strong resonance lines were roughly consistent with a wind with
$\dot{m}_{wind}$ of \EXPU{5}{-10}{\MSOL \: yr^{-1}}, or 10\% of
$\dot{m}_{disk}$, and a wind temperature of 45,000 K.  In their
model, the fractional abundance of an ion was assumed to be
constant throughout the wind. The fractional abundances of various
ions were then adjusted as part of a fitting procedure. They found
that the fractional ionization ranged from a few \% for C IV to
\EXPN{5}{-4} for O VI, roughly consistent with that expected by
photoionized material with a radiation temperature of
\EXPU{1.2}{5}{K}.

In an attempt to model all of the lines in the wind of a high
state disk in a self-consistent way, we have developed a new Monte
Carlo/Sobolev-based radiative transfer code -- Python.  Our
initial goal has been to be able to produce spectra that
qualitatively resemble those of disk-dominated CVs.  This is
useful, in and of itself, in order to help resolve which lines are
and which lines are not likely to be associated with wind
emission.  By comparison with observed spectra, we hope to use the
code to restrict the geometry and ionization structure of the
wind, and to improve estimates of basic wind parameters, including
the total mass loss rate in the wind.  By determining the
parameters of the wind, one can hope constrain the basic physics
of the wind, including, in principle, the driving mechanism.

\section{The Code}

Python is a Monte Carlo code based in large measure on the
techniques described in a series of papers by Leon Lucy and his
collaborators (Mazzali \& Lucy 1993; Lucy 1999 and references
therein). The geometry of the wind, by which we mean the density
and velocity structure of the wind, is specified in advance.  At
present, the geometries implemented include the CV wind
parameterizations described by Shlosman \& Vitello (1993) and by
Knigge et al.\ (1995), as well as the Caster \& Lamers (1979)
description of a spherical wind.  The latter is important for
comparisons with the many spectral simulations calculations that
have been made of O star winds.  Since we immediately project the
geometry of the wind onto a cylindrical grid, it is also
straightforward to calculate models for any reasonable gridded,
azimuthally symmetric flow.

There are four sources of radiation in the model: the disk, the
WD, the boundary layer and the wind itself.  The disk is assumed
to obey standard disk temperature and gravity relationships, and
the flux distribution from the disk and the star can be simulated
as blackbodies, Kurucz (1991) or Hubeny (1988) model atmospheres.
Linear limb darkening is assumed for all radiating surfaces. The
boundary layer is assumed to arise from the entire WD surface and
radiate isotropically; it is specified in terms of a blackbody
luminosity and temperature. Free-free, bound-bound, and free-bound
transitions, as well as electron scattering, comprise the physical
processes by which the wind radiates and absorbs.  At present, we
do not solve for the level populations self-consistently, but
treat each transition in a two-level approximation, allowing for
collisional de-excitation.

The astrophysical data required for the code are external to the
code itself. Our standard set of input data incorporates all
abundant elements from H through Ni and all ionization states with
ionization potentials of less than 500 eV. In the models
calculated here we have used the line list of Verner, Verner, \&
Ferland (1996), supplemented by a small number of UV transitions
with excited lower levels from  the Kurucz linelist.

The first step in a calculation of this sort is to determine the
ionization structure of the wind.  This is accomplished
iteratively.  An initial guess at the temperature and ionization
structure of the wind is made. Then a flight of typically 100,000
photon bundles is generated, representing the entire frequency
spectrum of all radiant sources. These photon bundles traverse the
wind, losing energy to it, until they emerge from the wind or
encounter the disk or WD.  As the photon bundle move, the
scattering optical depth $\tau_{scat}$ increases when the photon
frequency matches the Doppler-shifted resonant frequency of a line
and by electron scattering.  Photon bundles scatter when the
accumulated scattering optical depth
\begin{equation}
\tau_{scat} \ge -ln(1.-p),
\end{equation}
where p is a random number between 0 and 1 generated with a
uniform distribution. Because the optical depth at resonance can
be very large in the Sobolev approximation, escape probabilities
are used to determine how large a fraction of the photon bundle is
lost when it resonantly scatters.

As the photon bundles pass through the spatial grid, information
is accumulated to measure the effective radiation temperature, the
radiation density, and the energy absorbed by each element in the
grid. Based on these quantities, we make new estimates of the
ionization and radiation temperature. Specifically, we establish
the ionization structure based on a modified version of the ``on
the spot'' approximation:

\begin{equation}
\frac{n_{j+1} n_e}{n_j} =  W \left[ \xi + W (1-\xi) \right]
\left(\frac{T_e}{T_R}\right)^{1/2}
\left(\frac{n_{j+1}n_e}{n_j}\right)^*_{T_R}, \label{ionization}
\end{equation}
where the last (``*'') term on the right hand side of the equation
refers to abundances calculated using the Saha equation. In this
equation, $W$ is the effective dilution factor of the radiation
field, $\xi$ is the fraction of recombinations of an ion going
directly to the ground state, and $T_R$ and $T_e$ are the
radiation and electron temperature, respectively.

Following Mazzali \& Lucy (1993), we define the radiation
temperature $T_R$ by computing a $J_\nu$-weighted mean photon
frequency, $\overline{\nu}$, for each cell in each iteration. It
can be shown that in a blackbody radiation field described by
temperature $T_R$, this mean frequency satisfies $h
\overline{\nu}=3.832 k T_R$. Inverting this, we use
$\overline{\nu}$ to define $T_R$ via
\begin{equation}
T_R = \frac{h\overline{\nu}}{3.832 k}.
\end{equation}

Both geometric dilution and absorption will generally cause the
local radiation field, $J$, to have a value smaller than the
blackbody field at the local radiation temperature, $B(T_R)$. We
have dropped the subscript $\nu$ on $J$ and $B$ to denote that we
are concerned with frequency integrated quantities here. We
therefore define the effective dilution factor $W$ by demanding
that
\begin{equation}
J=W B(T_R)=W(\sigma / \pi) T_R^4.
\end{equation}

The electron temperature $T_e$ is adjusted so that power radiated
by the cell is equal to that absorbed.  This entire process is
then repeated until the structure relaxes to a stable
configuration (as measured by the changes in the temperature of
the grid points).  Typically, about 10 iterations are required.

Our reliance on this simple approximation is likely to be one of
the main weaknesses of the code at present.  It has the advantage
however that it does not require large numbers of photons for
convergence and that it approaches LTE in the LTE limit. We are
actively engaged in trying to determine the magnitude and nature
of the problems by comparison with publicly available codes. Our
initial assessment using Cloudy (Ferland 2000) does show some
differences, but the differences correspond to an effective
temperature offset of no more than a few thousand degrees under
conditions in which C IV and N V are expected in the wind.

Once the structure is converged, a second set of calculations is
carried out to generate a detailed spectrum at specific
inclination angles. In this case, the photon bundles are generated
only for the portion of the spectrum of specific interest. The
photons bundles are allowed to traverse the wind, and to scatter
and degrade in total luminosity as they do so. One could simply
sum up the weights of the photon bundles that escape within a
certain angle of the desired inclination to construct spectra.
However, it is actually possible to follow each photon bundle and
calculate at each interaction the relative probability of the
bundle escaping in the direction of one or more observers. By
using these ``extracted'' photons, one is able to produce spectra
with comparable photon statistics at any inclination angle. Since
the code also determines whether photon bundles encounter the
Roche surface of the secondary, it is also possible to follow the
spectra of highly inclined systems through eclipse.

\begin{figure}
\plotone{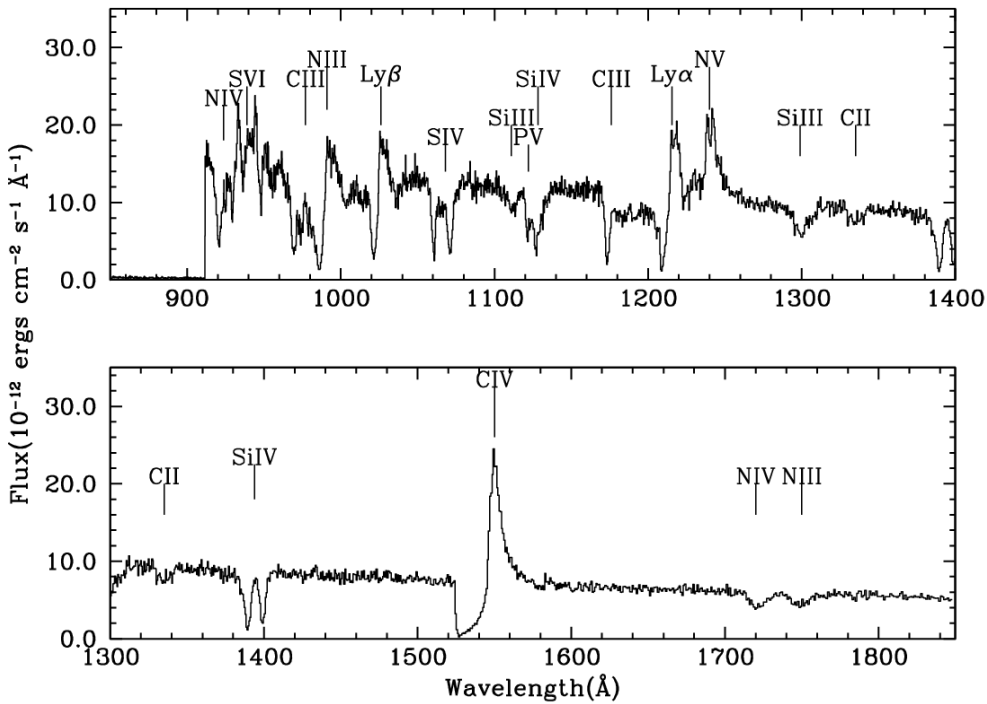} \caption{The calculated spectrum for the
fiducial Shlosman \& Vitello geometry as observed at an
inclination of 62.5\deg\ and assuming both the disk and the WD
radiate as blackbodies.}
\end{figure}

Fig.\ 1 illustrates the results of a typical simulation using
parameters that are identical to those of the fiducial model
discussed by Shlosman \& Vitello at an inclination angle of
62.5\deg. For this and other examples described below, $10^{6}$
photon bundles were created and followed through the wind to
create the final spectrum; the spectra cover 1000 \AA\ binned at
0.5 \AA\ intervals. The fiducial model is for a system with a
40,000 K WD with a mass of 0.8 \MSOL\ and radius of
\EXPU{6}{8}{cm}.  The disk is presumed to be in a steady state
with $\dot{m}_{disk}$ of \POW{-8}{\MSOL \: yr^{-1}} and to extend
from the WD surface and to 34 R$_{WD}$. The wind arises from the
disk between 4 and 12 R$_{WD}$, and at its inner edge has
streamlines that are at 12\deg\ with respect to the axis of the
system, while at the outer edge the streamlines are 65\deg\ with
respect to the normal.  The scale length for defining the velocity
of the wind is 100 R$_{WD}$ so the wind does not reach terminal
velocity until the wind is well outside the system. The terminal
velocity along each stream line is 3 times the escape velocity at
the footpoint of the wind, so that the most rapidly moving
portions of the wind are located at its inner edge. For a more
complete description of the kinematic model, see Shlosman \&
Vitello (1993).

In the simulated spectrum shown in Fig.\ 1, both disk and the WD
spectral distributions have been formed from appropriately
weighted blackbodies, so that all of the features in the spectrum
are due to the effects of the wind, including the sharp cutoff at
912 \AA. In this simulation, C IV shows a well-developed P Cygni
profile. Ly$\alpha$ and Ly$\beta$ also show emission wings to
their profiles, even though the ionization fraction of H in the
outer wind is quite low.  O VI is not evident.  A number of lower
ionization potential lines, such as the Si IV doublet at 1400 \AA\
appear only in absorption. These ions are present only near the
disk and as a result we see them only in absorption along the line
of sight to the disk.

\begin{figure}
\plotone{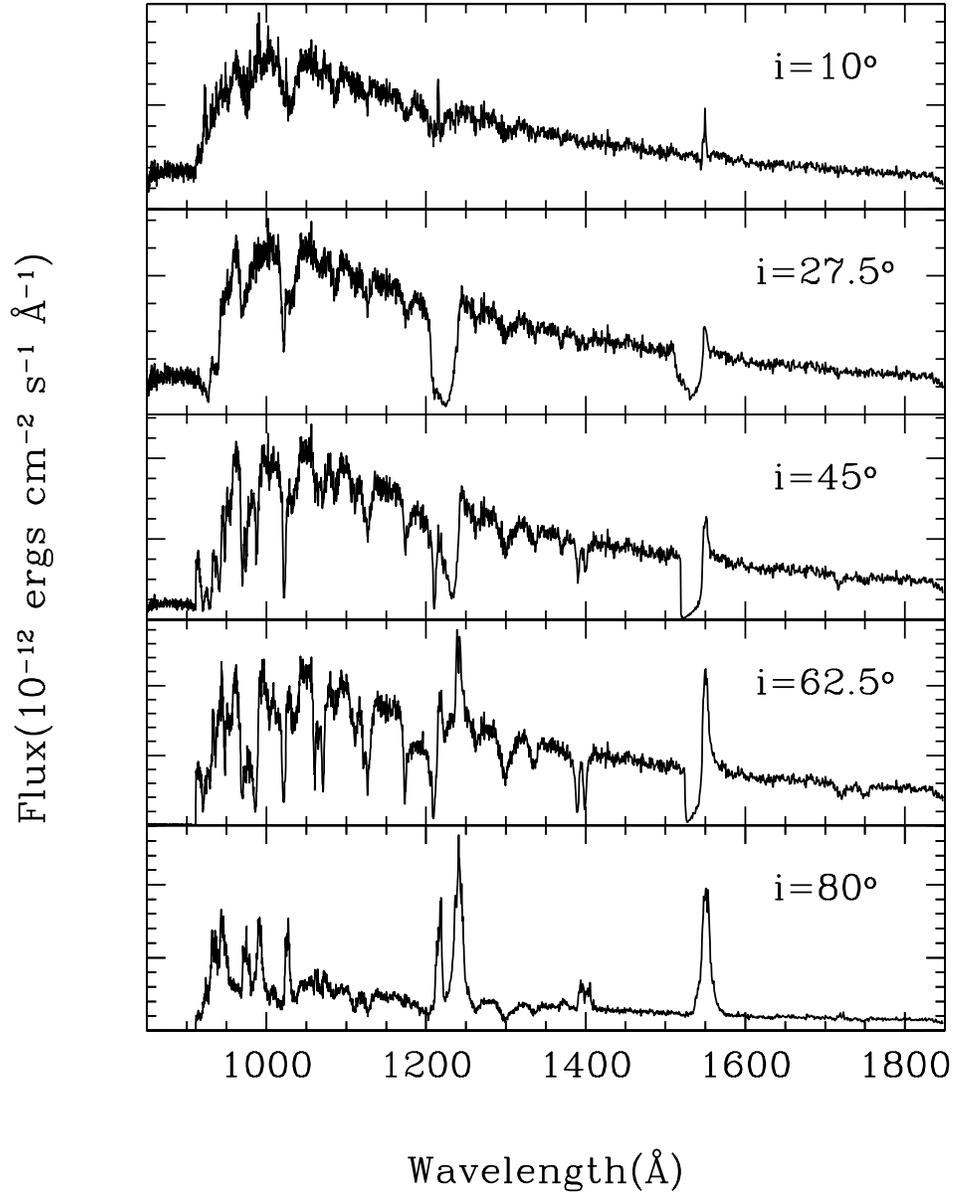} \caption{The calculated spectrum for the
fiducial Shlosman \& Vitello geometry, which has $\dot{m}_{disk}$
of \POW{-8}{\MSOL \: yr^{-1}} and $\dot{m}_{wind}$ of
\POW{-9}{\MSOL \: yr^{-1}} as a function of inclination angle. In
contrast to Fig.\ 1, here the disk and WD were simulated in terms
of appropriately weighted Hubeny (1988) model atmospheres.}
\end{figure}

Our C~IV profiles differ in detail from those produced by Shlosman
\& Vitello (1993). In particular we find somewhat wider absorption
profiles at moderate inclination angles and somewhat greater
emission at high inclination.  This is almost surely due to
differences in the ionization structure in the two calculations.
In the Shlosman \& Vitello model calculation, the dominant state
of C far from the WD appears to be C~V, while our wind retains
significant C~IV at large radii. It is not clear at this stage
which ionization structure is more correct.  The implication, if
our ionization calculation is more accurate, is that we will
require somewhat lower mass loss rates than did Shlosman \&
Vitello for modelling the spectra of specific CVs.

In Fig.\ 2, the same model is shown for inclinations ranging from
10\deg\ to 80\deg, but in this case the disk and WD are simulated
in terms of appropriately weighted stellar atmospheres. At 10\deg,
when the observer views the system from ``inside'' the inner wind
cone, the effects of the wind on the spectrum are quite subtle,
and nearly all of the features are photospheric with the exception
of the narrow emission profiles of Ly$\alpha$ and C IV.  At angles
of 27.5, 45, and 62.5\deg, the observer is viewing the disk
through larger and larger columns of wind material and more ions
become apparent. At 62.5\deg, the line of sight skims over the
outer lower temperature portion of the disk and therefore more and
more low ionization lines appear in the spectrum.  Emission
becomes more and more important in ions that are extended in the
wind, such as C IV and N V, not so much because the flux due to
emission has increased, but primarily because the projected size
and hence emission from the disk is decreasing.  A comparison of
the spectrum in Fig.\ 1 to the 62.5\deg\ panel in Fig.\ 2. shows
how important an accurate ``photospheric'' model of the disk and
WD remains in the FUV. Finally, at 80\deg, when the observer views
the system from ``outside'' the wind cone, the spectrum appears
dominated by emission lines.

\begin{figure}
\plotone{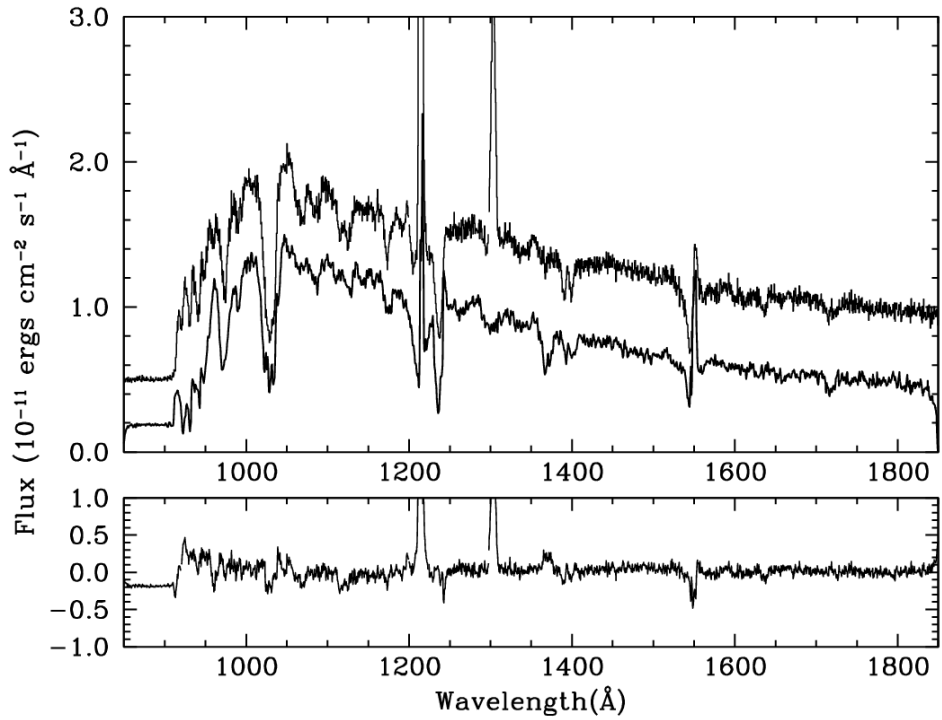} \caption{A comparison between the HUT
spectrum of Z Cam and a model spectrum calculated using a
prescription for the wind geometry similar to that suggested by
Knigge et al.\ (1997). The model has been smoothed slightly to
reflect the resolution of HUT. For clarity, the data have been
displaced above the model in the top panel. The strong features at
1216 \AA\ and 1304 \AA\ are air glow lines. The difference between
the data and the model is shown in the bottom panel. }
\end{figure}

\section{Comparisons to Spectra of Real CVs}

In the absence of a physical model of the wind from a dwarf nova
in outburst, any description of a bi-conical flow necessarily has
a large number of parameters. It is far from clear exactly how to
explore the parameter space, especially since the nature of the
boundary layer is unknown in most if not all of the systems
involved.  Furthermore, given our limited experience at this
point, it is not obvious whether more than one wind structure can
produce a similar wind spectrum.

However, an example of the type of ``fit'' that can be obtained is
shown in Fig.\ 3, which shows data obtained with HUT during a
normal outburst of Z Cam compared to a model spectrum generated
using a wind geometry nearly identical to that described by Knigge
et al. (1997).  Specifically, the model shown here has the same
amount of collimation, the same acceleration length and terminal
velocity, as well as  the same value of \EXPU{5}{-9}{\MSOL} for
$\dot{m}_{disk}$ (originally derived from fitting the continuum
spectrum). The model shown is for an inclination of 49\deg, which
is in the range allowed for Z Cam . Qualitatively, the model
spectrum and the data resemble one another quite well. The strong
resonance lines of C IV, N V, and O VI have approximately the
right profiles, and many of the features of the lower ionization
lines can be seen in both the model and in the data. The main
difference between the two models is in $\dot{m}_{wind}$. Knigge
et al.\ (1997) assumed that $\dot{m}_{wind}$ was
\EXPU{5}{-10}{\MSOL} and found that ionization fractions of C IV,
NV, and O VI were relatively low, a few \% or less.  We however
find a better fit with $\dot{m}_{wind}$ of \EXPU{1}{-10}{\MSOL \:
yr^{-1}}.  The ionization fractions of C IV, N V and O VI are
calculated in Python and turn out to be substantial in much of the
wind.

\section{Summary}

The inherent advantages of a Monte Carlo approach in simulating
the spectra of cataclysmic variables is the ease with which one
can model complex geometries.  In the Monte Carlo approach,
axially symmetric winds are not significantly more difficult than
spherical winds.  In the case of Python, we have now progressed to
the point of spectral verisimilitude, the point where it is
difficult to distinguish a simulation from data.  The wind
prescriptions required to produce verisimilitude are not that
different from those which were developed to model C IV line
profiles.

On the other hand, preliminary comparisons of the models and data
suggest that considerable work will be required to reproduce the
spectra of some dwarf novae.  Although we are having some
successes, such as the model shown here for Z Cam and with the
FUSE spectrum of SS Cygni (Froning et al.\ 2001), it is quite
difficult to reproduce in detail the observed spectra over a broad
wavelength range of other CVs, far more difficult than to
reproduce an individual line profile.  It is not yet clear whether
this is due to the fact that we have yet to hit upon the correct
parameters in the geometries implemented in Python, or whether
more sophisticated treatment of the physics involved in a wind is
involved.

To address the former, we are embarking upon an effort to fully
explore the parameter space inherent in the existing models and to
compare these models to a variety of HUT, FUSE and HST spectra of
high state CVs. This should allow us to determine whether there is
a single class of models, broadly or narrowly collimated for
example, that can be used to approximate the wind geometry in
majority of CVs. To address the latter, we are beginning
comparisons with models of O star winds for which there is
considerable history and expertise. This should provide a clear
validation of the code itself and suggest areas where the physics
must be improved.

\acknowledgements{This work was supported by NASA through grant
G0-7362 and GO-8279 from the Space Telescope Science Institute,
which is operated by AURA, Inc., under NASA contract NAS5-26555.}

\end{document}